\documentclass[twocolumn,floats,showpacs,prb]{revtex4}
\usepackage[above]{placeins}
\usepackage{amsmath}
\usepackage{graphicx}

\begin{document}

\title{ Interpretation of high-pressure experiments on FeAs superconductors}
\author{Xiuqing Huang$^{1,2}$}
\email{xqhuang@nju.edu.cn}
\affiliation{$^1$Department of Physics and National Laboratory of Solid State
Microstructure, Nanjing University, Nanjing 210093, China \\
$^{2}$ Department of Telecommunications Engineering ICE, PLAUST, Nanjing
210016, China}
\date{\today}

\begin{abstract}
In two recent articles (cond-mat/0606177 and arXiv:0804.1615), we have
suggested a unified theory of superconductivity\ based on the real-space
spin-parallel electron pairing and superconducting mechanism and have shown
that the stable hexagonal and tetragonal vortex lattices (the optimal doping
phases) can be expected in the newly discovered LaO$_{1-x}$F$_{x}$FeAs ($%
x_{0}=1/7\approx 0.1428$) and SmO$_{1-x}$F$_{x}$FeAs ($x_{0}=1/6\approx
0.1667$), respectively. In this paper, we present a theoretical study of the
effects of hydrostatic and anisotropic pressure on the superconducting
transition temperature $T_{c}$ of the Fe-based layered superconductors based
on the above mentioned theory. Our results indicate a strong
doping-dependent pressure effects on the $T_{c}$ of this compound system.
Under high hydrostatic pressure, we find that $dT_{c}/dP$ is negative when $%
x>x_{0}$ (the so-called overdoped region) and is positive when $x<x_{0}$
(the so-called underdoped region). Qualitatively, our finding is in good
agreement with the existing experimental data in LaO$_{1-x}$F$_{x}$FeAs ($%
x=0.11<1/7$) (arXiv:0803.4266) and SmO$_{1-x}$F$_{x}$FeAs ($x=0.13<1/6$ and $%
x=0.3>1/6$) (arXiv:0804.1582). Furthermore, $T_{c}$ of both overdoped and
underdoped samples shows an increase with uniaxial pressure in the charge
stripe direction and a decrease with pressure in the direction perpendicular
to the stripes. We suggest that the mechanism responsible for the pressure
effect is not specific to the iron-based family and it may also be
applicable to other superconducting materials.
\end{abstract}

\maketitle

\section{Introduction}

The surprising discovery of superconductivity in FeAs superconductors \cite%
{kamihara} has stimulated intense interest on the mechanism for
superconductivity in this new high-$T_{c}$ superconductor family.$^{2-23}$
It is well known that element substitution is often served as an effective
method to raise $T_{c}$ due to the possible internal pressure induced by the
replacement of the smaller elements.$^{24-30}$ In this way, physicists
around the world have pushed the transition temperature of the new
superconductors from 26 $K$ to 55 $K$. Moreover, it is an experimental fact
that the $T_{c}$ of superconductors can also be tuned by the external high
pressure. Theoretically, the possibility of enhancing $T_{c}$ has been
suggested in this class of compounds by applying pressure.$^{31-35}$
Experimentally, it was found that the $T_{c}$ of LaO$_{1-x}$F$_{x}$FeAs ($%
x=0.11$) increases almost linearly (at a rate of $1.2$ K/GPa) with the
increasing of hydrostatic pressure, \cite{wlu} in favor of the previous
expectations. The external pressure effects were also reported on SmO$_{1-x}$%
F$_{x}$FeAs, Lorenz \textit{et al}. \cite{lorenz} found that in the $x=0.13$
sample $T_{c}$ increases under hydrostatic pressure with the rate $%
dT_{c}/dP\simeq 0.9$ K/GPa. However, the $T_{c}$ of the $x=0.3$ sample is
suppressed instead by pressure at a rate of $dT_{c}/dP\simeq -$ 2.3 K/GPa,
in contrast to the theoretical predictions. These measurements reveal that
the pressure variation of $T_{c}$ of FeAs compounds may be strongly
dependent on the doping level of $x$.

In the earlier works, \cite{huang1,huang2} we have proposed a real space
mechanism of high-$T_{c}$ superconductivity which can naturally explain the
complicated problems, such as pairing mechanism, pairing symmetry, charge
stripes, optimal doping, magic doping fractions, vortex structure, phase
diagram, Hall effect, etc.\cite{huang1,huang2} Based on the mechanism, the
relationship between the superconducting vortex phases and the optimal
doping levels of FeAs superconductors were analytically given. We predicted
that the optimal doping levels are $x=1/7\approx 0.1428$ (LaO$_{1-x}$F$_{x}$%
FeAs and La$_{1-x}$Sr$_{x}$OFeAa) and $x=1/6\approx 0.1667$ (Ce$_{1-x}$O$%
_{x} $FFeAs, SmO$_{1-x}$F$_{x}$FeAs, PrO$_{1-x}$F$_{x}$FeAs and CdO$_{1-x}$F$%
_{x}$FeAs) which are found to be in excellent agreement with the
experimental data (LaO$_{1-x}$E$_{x}$FeAs: $x=0.12$, La$_{1-x}$Sr$_{x}$%
OFeAa: $x=0.13$, Ce$_{1-x}$O$_{x}$FFeAs: $x=0.16$, SmO$_{1-x}$F$_{x}$FeAs: $%
x=0.15$, PrO$_{1-x}$F$_{x}$FeAs: $x=0.16$, and CdO$_{1-x}$F$_{x}$FeAs: $%
x=0.17$), furthermore, it is shown that when $x=1/7$ the triangular array of
vortices can be expected in the related samples, while $x=1/6,$ the
corresponding vortex lattice structures have a tetragonal symmetry. \cite%
{huang0} Moreover, although the new layered materials resemble the cuprates
in some ways, it was shown that the new compounds belong to non-pseudogap
superconductors.

In the present paper, we try to extend the application of the theory to the
effects of pressure on the superconducting properties of FeAs
superconductors. Furthermore, we aim to explore a universal relationship
among the pressure effects (hydrostatic and anisotropic) on the transition
temperature $T_{c}$, lattice constants, vortex lattices, charge (magnetic)
stripes and the doping levels in the superconductors.

\begin{figure*}[tbp]
\begin{center}
\resizebox{1.5\columnwidth}{!}{
\includegraphics{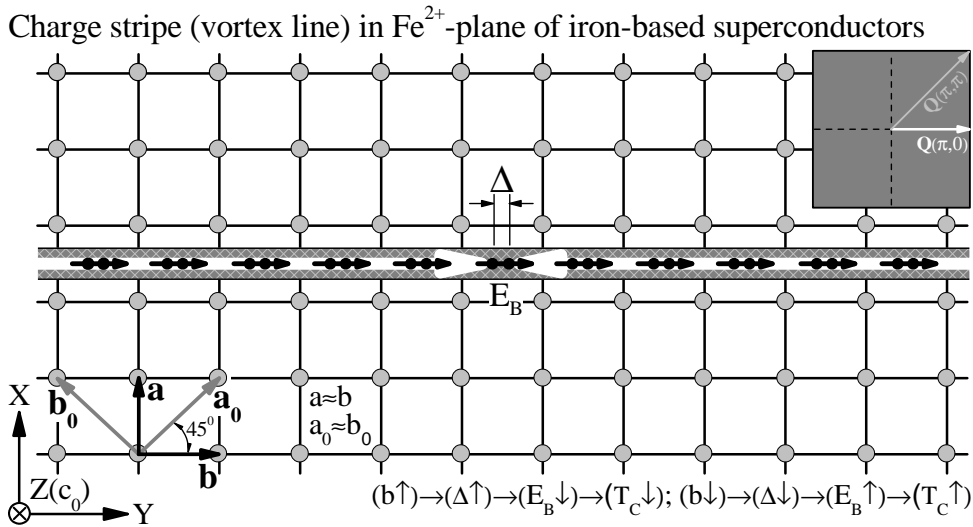}}
\end{center}
\caption{{}The real space collective confinement picture (vortex line) in
the superconducting iron plane of FeAs superconductors. The superconducting
transition temperature $T_{c}$ is uniquely determined by lattice constant $b$%
. When the superconducting vortex phase with ordering wave-vector $\mathbf{Q}%
=(\protect\pi ,0)$, the spin-density-wave (SDW) is suppressed. While the
ordering of $\mathbf{Q}=(\protect\pi ,\protect\pi )$, an intensive SDW is
inspired along the charge stripe (vortex line) and the superconductivity is
totally suppressed. This implies the FeAs family is possible in $d$-wave
superconductors.}
\label{fig1}
\end{figure*}

\section{How to raise the superconducting transition temperature?}

To raise the superconducting transition temperature up to the room
temperature is still a dream of scientists today. Although almost 100 years
have passed since the first discovery of superconductivity in mercury in
1911 by H. Kamerlingh Onnes, \cite{onnes} disappointedly, so far scientists
cannot draw a convincing physical picture: What is the superconducting
phase? Undoubtedly, if this situation persists, it is never possible for
scientists to achieve a reasonable understanding of the mysterious
superconducting phenomenon and the dream of superconductivity at higher
temperatures (perhaps even room temperature) will always remain as a dream.
In other words, we need to establish the nature of the superconducting
charge carriers before thinking about how to enhance $T_{c}$. Here, we would
like to point out that the following three factors play a central role in
raising the $T_{c}$ of superconductors.

\subsection{Charge and magnetic stripes (vortex lines)}

\begin{figure}[bp]
\begin{center}
\resizebox{1\columnwidth}{!}{
\includegraphics{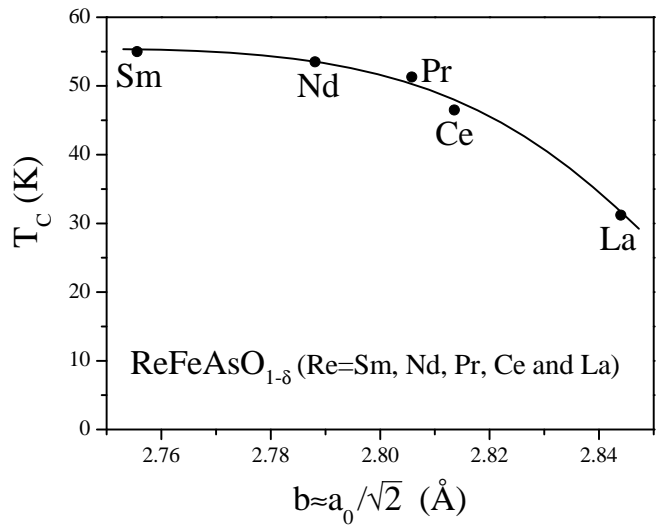}}
\end{center}
\caption{{}The relationship between the lattice constant $b$ (or the
distance of the nearest-neighbor Cooper pairs) and $T_{c}$ in FeAs
superconductors.}
\label{fig2}
\end{figure}

First, to exhibit superconductivity, the cooper pairs should be condensed
into a charge river (stripe) and a stable quasi-one-dimensional
\textquotedblleft freeways\textquotedblright\ should be built naturally in
the superconductors. In the previous study, the real space collective
confinement pictures have been introduced into the conventional and cuprate
superconductors. Figure \ref{fig1} shows the real space collective
confinement picture in the superconducting iron plane of FeAs
superconductors. In this case, a real space long range magnetic order (spin
parallel) and superconductivity coexist to form a dimerized charge
supersolid (a charge-Peierls dimerized transition), as seen in Fig. \ref%
{fig1}. Note that the 3D coordinate system with axis X$(a)$, Y$(b)$, Z$(c)$,
which is $45^{0}$ rotated along the $c_{0}$ direction of the realistic
crystal axis $a_{0}$, $b_{0}$ and $c_{0}$. Because of the strong
interactions among the cooper pairs, the lattice constant $b$ is usually
slightly larger than $a$. Hence, the elementary square plaquette become
slightly rectangular. But for the simplification of discussion, we assume
that $a=b$ and $a_{0}=b_{0}.$ Under such an approximation, the charge and
magnetic order (vortex line) with ordering vector $\mathbf{Q}=(\pi ,0)$
appears to be compatible with superconductivity in the system, as shown in
Fig. \ref{fig1}. When temperature $T\neq 0$, the so-called spin density wave
(SDW), a competing state against the superconductivity, will be inspired in
the metallic charge stripe (vortex line) due to thermal fluctuations and the
superconductivity and SDW can coexist along this stripe. When temperature $%
T=0,$ the SDW order is totally suppressed and the corresponding stripe is
referred to as the superconducting ground state.

From Fig. \ref{fig1} it is apparent that, for the given lattice parameters,
the superconducting phase of $\mathbf{Q}=(\pi ,0)$, or $\mathbf{Q}=(0,\pi )$%
, has the highest superconducting transition temperature due to the most
intensive confinement effect. While the phase with ordering wave-vector $%
\mathbf{Q}=(\pi ,\pi )$ corresponds to the SDW phase where the
superconductivity is totally suppressed. This implies the FeAs family is
possible in $d$-wave superconductors. Furthermore, for the best
superconducting order of $\mathbf{Q}=(\pi ,0)$ presented in Fig. \ref{fig1},
the distance $\Delta $ between two electrons of one cooper pair decreases
with the decreasing of the lattice constant $b$, as a consequence, increase
the pair binding energy $E_{B}$ and the superconducting transition
temperature $T_{c}$. This conclusion has been well confirmed by the recent
chemical (element) substitution experiments in FeAs superconductors, \cite%
{zaren} as shown in Fig. \ref{fig2}, the shrinking of crystal lattice can
effectively enhance the superconducting transition temperature.

\subsection{Vortex lattices}

\begin{figure}[tbp]
\begin{center}
\resizebox{0.85\columnwidth}{!}{
\includegraphics{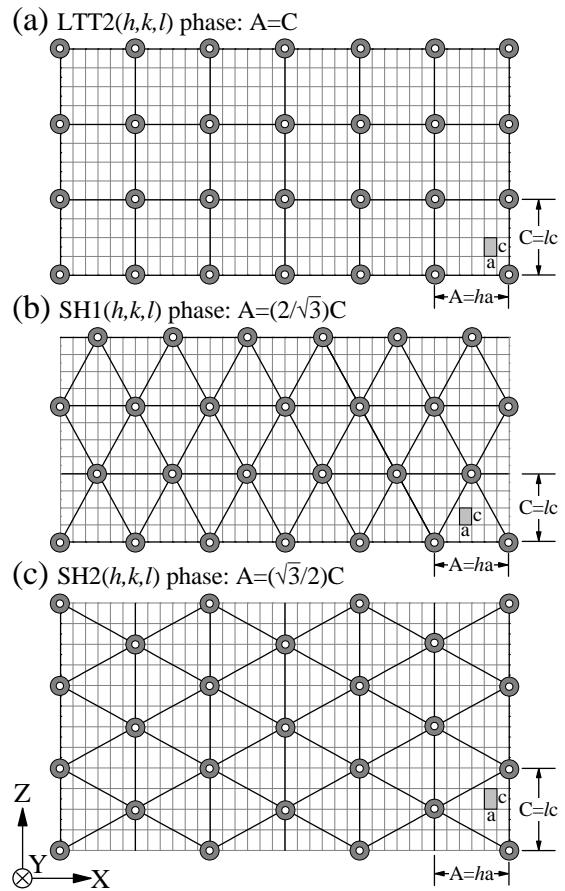}}
\end{center}
\caption{{}{} Three possible stable superconducting charge-stripe orders
(vortex lattices) in the XZ plane perpendicular to the plane of CuO$_{2}.$
(a) The LTT2($h,k,l$) phase, here the tetragon superlattice structure ($A=C$%
) is shown$,$\ (b) the trigonal SH1($h,k,l$) phase with $A/C=2/\protect\sqrt{%
3}$, and (c) the trigonal SH2($h,k,l$) phase where $A/C=\protect\sqrt{3}/2.$}
\label{fig3}
\end{figure}

Second, to maintain a stable and durable superconducting phase, the metallic
charge stripes of Fig. \ref{fig1} (vortex lines) should self-organize into a
`superlattice' (vortex lattices) with the primitive cell $(A,B,C)=(ha,kb,lc)$%
. It has been argued that the physically significant critical value for the
most stable vortex lattice is that at which $T_{c}$ is maximum. In this
sense, the LTT2 and the simple hexagonal (SH) phases (vortex lattices) might
be the ideal candidates for the stable charge-stripe order of paired
electrons. In the LTT2($h,k,l$) phase, as shown in Fig. \ref{fig3} (a), the
charge stripes have a tetragonal symmetry in XZ plane in which the
superlattice constants satisfy
\begin{equation}
\frac{A}{C}=\frac{ha}{lc}=1.
\end{equation}%
While in simple hexagonal (SH) phases, as shown in Figs. \ref{fig3} (b) and
(c), the charge stripes possess identical trigonal crystal structures. In
the SH1($h,k,l$) phase [see Fig. \ref{fig3} (b)], the superlattice constants
have the following relation
\begin{equation}
\frac{A}{C}=\frac{ha}{lc}=\frac{2\sqrt{3}}{3}\approx 1.154700.
\end{equation}%
For the SH2($h,k,l$) phase of Fig. \ref{fig3} (c), this relation is given by
\begin{equation}
\frac{A}{C}=\frac{ha}{lc}=\frac{\sqrt{3}}{2}\approx 0.866025.
\end{equation}

We have shown that the appearance of the SH (or LTT2) vortex lattice\ is a
common feature of the optimally doped superconductors. But, for non-optimal
doping we found that the vortex lattices tend to form the superconducting
low-temperature orthorhombic (LTO) phase where the superlattice constants
satisfy $A\neq B\neq C.$

\subsection{Stripe-stripe interaction}

\begin{figure}[tbp]
\begin{center}
\resizebox{1\columnwidth}{!}{
\includegraphics{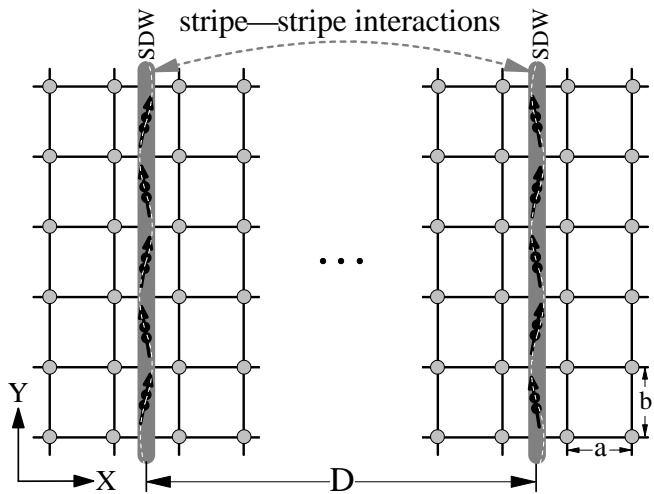}}
\end{center}
\caption{{}Stripe-stripe interaction inside the vortex lattice of FeAs
superconductors. An appropriate (or optimal) stripe-stripe distance $D$ (not
too close, not too far) is helpful for a higher $T_{c}$.}
\label{fig4}
\end{figure}

Third, the formation of stripe patterns (vortex lattices) is generally
attributed to the competition between short-range attractive forces and
long-range repulsive forces. \cite{seul} Therefore, it is inevitable that
there exist the intrinsic stripe-stripe interactions among the vortex lines
inside the vortex lattices, as illustrated in Fig. \ref{fig4}. On the one
hand, the interactions among stripes (vortex lines) are necessary for the
establishment of vortex lattices of Fig. \ref{fig3}. On the other hand, the
stripe-stripe interactions may induce the collective spin density wave (SDW)
excitations (harmful for superconductivity) along the vortex lines. With
increasing doping concentration in the system, stripe-stripe interactions
become more important. Obviously, an appropriate (or optimal) stripe-stripe
distance $D$ (not too close, not too far) is helpful for a higher $T_{c}$
(see Fig. \ref{fig4}). Charge carrier doping has been proved to be the best
and convenient way to control the stripe-stripe distance and interaction in
the doped superconductors. In our opinion, the optimally doping means that
the most stable vortex lattice (with an optimal stripe-stripe distance) has
been successfully established in the superconducting sample and the SDW
state has been greatly suppressed inside the vortex lattice. For the LTO($%
h,k,l$) superconducting vortex phase, two stripe-stripe distances $D_{xy}$
and $D_{z}$ are given by%
\begin{equation}
D_{xy}=ha,\quad D_{z}=lc.  \label{distant}
\end{equation}

The main results obtained so far and discussed in this section already allow
one to draw some useful conclusions about the relationship between vortex
structure and $T_{c}$. It is shown that to effectively raise the
superconducting transition temperature, the following three conditions
should be paid attention: (i) a compact one-dimensional charge-magnetic
stripe (vortex line); (i) a stable vortex lattice structure; and (iii) an
adequate stripe-stripe spacing and interaction.

\section{High pressure effects}

Shortly after the discovery of the superconductivity in LaO$_{1-x}$F$_{x}$%
FeAs, one of the central concerns of high-temperature iron-based
superconductors is how to raise the superconducting transition temperature.
Apart from the great efforts to the effect of element substitution on the $%
T_{c}$, the hydrostatic pressure induced $T_{c}$ increasing effects have
been reported in this class of compounds. \cite{wlu,lorenz,takahashi} While
these interesting experimental results seem to gain important insight into
FeAs superconductors, we find that there is no theory that can be applied to
explain them.

\begin{figure}[tbp]
\begin{center}
\resizebox{0.95\columnwidth}{!}{
\includegraphics{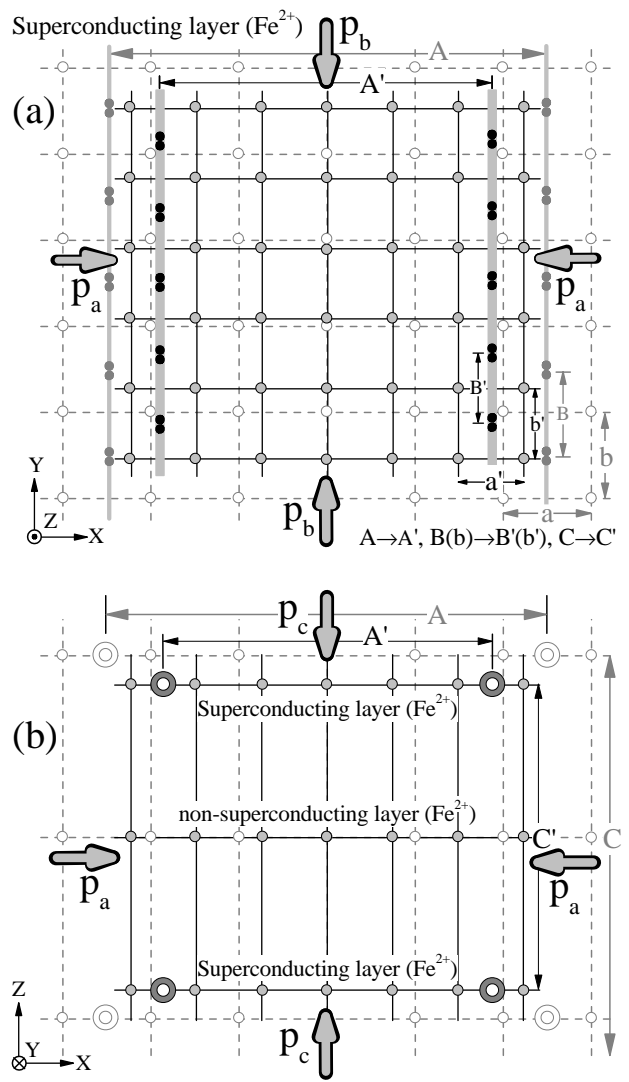}}
\end{center}
\caption{{}The hydrostatic pressure effects on the FeAs superconductors. The
pressure-induced shrinkage of the vortex lattice, (a) in the superconducting
Fe plane, (b) in the plane perpendicular to the vortex lines.}
\label{fig5}
\end{figure}

It is clear that a promising theory of superconductivity should explain
these results in addition to other properties. Here we show how the recently
suggested superconductivity theory can explain the pressure dependence of $%
T_{c}$ in iron-based superconductors. To the best of our knowledge, this is
the first theoretical attempt to do so. Based on the real-space
spin-parallel electron pairing and superconducting mechanism, the effects of
pressure can be visually illustrated in Fig. \ref{fig5} of LTO
superconducting vortex lattice in FeAs superconductors. One can see clearly
that the direct effect of pressure is to shrink the lattice constants, as a
result, changing the structure of vortex lattice. Hence, the pressure
effects can be described simply by
\begin{eqnarray*}
A( &=&ha)\rightarrow A^{\prime }(=ha^{\prime }), \\
B( &=&kb)\rightarrow B^{\prime }(=kb^{\prime }),k=1, \\
C( &=&lc)\rightarrow C^{\prime }(=lc^{\prime }),
\end{eqnarray*}%
where ($A,B,C$) and ($A^{\prime },B^{\prime },C^{\prime }$) are the
superlattice constants without and under the pressure. In the following
sections, we shall focus our attention on the problem: how can the pressure
affect the $T_{c}$ merely by varying the lattice (or superlattice) constants
of the superconductors?

\subsection{ Pressure effects on a single vortex line}

\begin{figure}[bp]
\begin{center}
\resizebox{1\columnwidth}{!}{
\includegraphics{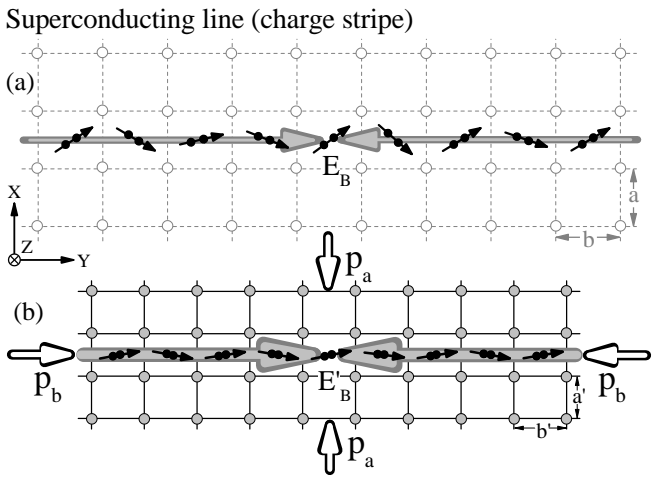}}
\end{center}
\caption{Pressure effects on a single vortex line. (a) Without pressure,
where 1D magnetic vortex line may coexist with a weak spin density wave
(spin fluctuation) with the pair binding energy $E_{B}$. (b) The external
pressure leads to the decreasing of lattice constants ($b^{\prime }<b)$ and
the suppression of the spin fluctuation, consequently, increase the binding
energy ($E_{B}^{\prime }>E_{B}$) and $T_{c}$.}
\label{fig6}
\end{figure}

Figure \ref{fig6}(a) shows a quasi-one-dimensional charge-magnetic
superconducting stripe (vortex line) where the stripe-stripe interactions do
not exist. In this ideal system, the values of lattice constant ($%
a\rightarrow a^{\prime }$ and $b\rightarrow b^{\prime }$) are decreasing
monotonously with the increasing external pressure, as shown in Fig. \ref%
{fig6}(b). According to the above discussions (see Fig. \ref{fig1}), we can
see that the pressure leads to the decreasing of the lattice constants,
consequently, increase the pair binding energy $E_{B}$ and narrow and
eventually eliminate the magnetic excitations in spin density wave (SDW)
state. Hence, it is expected that the pressure dependence of $dT_{c}/dT$ is
always positive in the quasi-one-dimensional vortex line. In other words, $%
T_{c}$ will be found to increase almost linearly upon increasing external
pressure in this specific system.

\begin{figure}[tbp]
\begin{center}
\resizebox{0.9\columnwidth}{!}{
\includegraphics{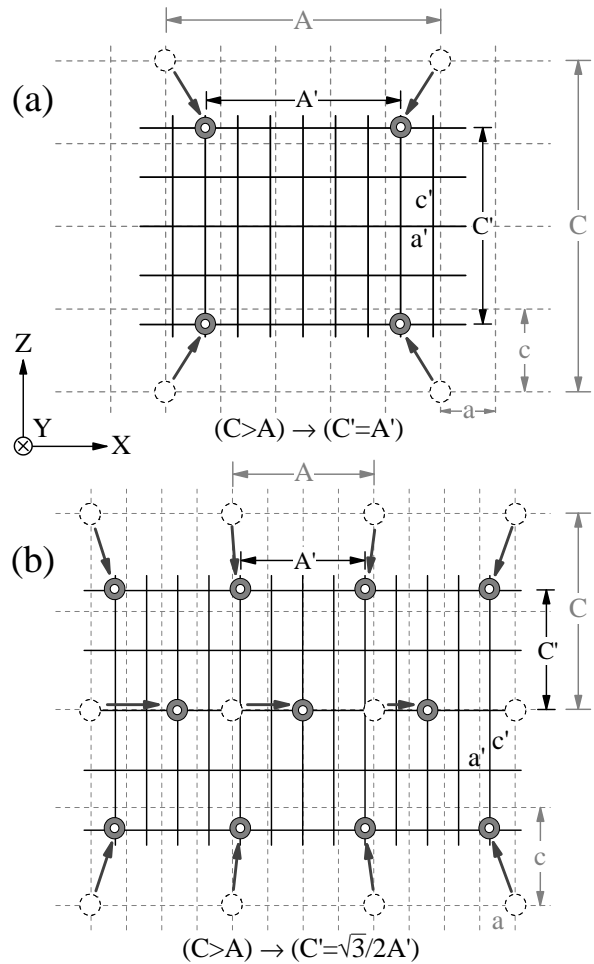}}
\end{center}
\caption{Two possible pressure-induced vortex lattice phase transitions in
FeAs superconductors. (a) LTO to LTT2, (b) LTO to SH1.}
\label{fig7}
\end{figure}

\subsection{Structural phase transition in the vortex lattice}

The superconducting LTT2, SH1 and SH2 (see Fig. \ref{fig3}) vortex phases,
as discussed above, are much more stable than the LTO superconducting phase.
As a consequence, the highest $T_{c}$ phase is usually related to LTT2, SH1
or SH2 vortex lattice. Figure \ref{fig7} shows two possible pressure-induced
vortex lattice phase transitions in FeAs superconductors. Shown in Fig. \ref%
{fig7}(a) is the LTO-LTT2 structural phase transition, while an example of
LTO-SH1 vortex lattice phase transition is illustrated in Fig. \ref{fig7}%
(b). These suggest that pressure can play an important role on pushing low $%
T_{c}$ superconducting vortex phase toward the main (optimal)
superconducting phase.

To depict the difficulty level of vortex lattice phase transition (or
compressibility) in superconductors, one can define the following
\textquotedblleft stiffness criterion\textquotedblright\ for the underdoped
superconducting LTO($h,k,l$) vortex phase

\begin{equation}
\Theta =\kappa \left[ \left( \frac{1}{hkl}\right) _{0}-\frac{1}{hkl}\right]
\frac{1}{abc}=\kappa \frac{(x_{0}-x)}{2abc},  \label{stiff}
\end{equation}%
where $\kappa $ is a material-related constant, $x$ is the doping level in
the sample. And $x_{0}$ and $(1/hkl)_{0}$ are the optimal doping level and
the corresponding vortex lattice index, respectively.

The relation of Eq. (\ref{stiff}) implies that the stiffness of
superconductor is direct proportion to the doping level $x$, but is inverse
proportion to the unit-cell volume ($abc$). The smaller the parameter $%
\Theta $, i.e., the softer the corresponding compound system, the larger the
$dT_{c}/dP$ value is, and vice versa. Thus, we have
\begin{equation}
\frac{dT_{c}}{dP}\propto \frac{1}{\Theta ^{\alpha }},  \label{temperature}
\end{equation}%
where $\alpha $ is a positive constant.

From Eq. (\ref{stiff}) and (\ref{temperature}), it is clear that $dT_{c}/dP$
value depends strongly on the doping level and the lattice constants. We
find that pressure can either promote or suppress the superconducting $T_{c}$%
, depending on the doping level of $x$. For the cuprate superconductors, it
has been shown that the $T_{c}$ varies with carrier concentration $n$
following a universal parabolic rule with $T_{c}$ peaks at a carrier
concentration $n_{o}$. \cite{cwchu} Later, it has also been proved that $%
dT_{c}/dP$ is negative when $n>n_{o}$ and positive when $n<n_{o}$. \cite%
{presland} Obviously, the above two expressions give a reasonable agreement
with the results in the cuprate superconductors. This consistency implies
that the Eq. (\ref{stiff}) and (\ref{temperature}) are not specific to the
iron-based family and it may also be applicable to other doped
superconducting materials.

\begin{figure}[tbp]
\begin{center}
\resizebox{1\columnwidth}{!}{
\includegraphics{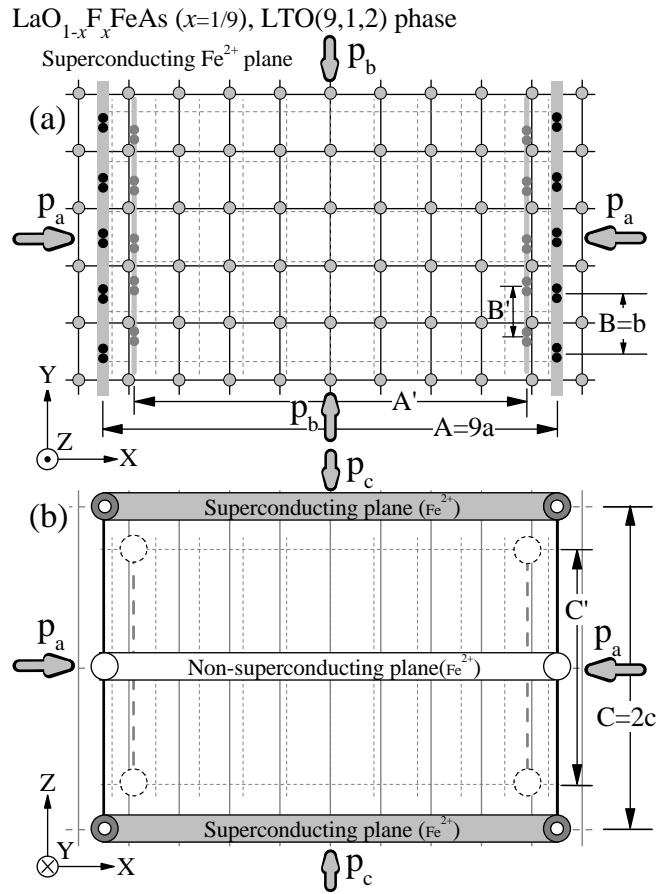}}
\end{center}
\caption{The schematic plot of the pressure effects on LTO($9,1,2$)
superconducting phase of the LaO$_{1-x}$F$_{x}$FeAs with $x=1/9\approx 0.111$%
. The pressure-induced distortion of vortex lattice, (a) the doped Fe plane,
(b) the vortex lattice plane. }
\label{fig8}
\end{figure}

\subsection{LaO$_{1-x}$F$_{x}$FeAs}

In LaO$_{1-x}$F$_{x}$FeAs, based on the experimental lattice constants ($a=%
\sqrt{2}a_{0}/2=2.85\mathring{A}$ and $c=c_{0}=8.739\mathring{A}$), we have
predicted that the optimal doping levels is $x_{0}=1/7\approx 0.1428$ with
the hexagonal vortex SH1($7,1,2$) lattice having the stable trigonal
structure. \cite{huang0} The hydrostatic-pressure effects on the
superconducting transition temperature ($T_{c}$) of the LaO$_{1-x}$F$_{x}$%
FeAs ($x=0.11$) have been recently reported by two research groups. \cite%
{wlu,takahashi} These results corroborate the suggested external
pressure-induced $T_{c}$-enhancement in the compound. It should be pointed
out that the $x=0.11$ sample lie in the underdoped region, in favor of the
positive pressure effect on $T_{c}$.

We note that $x=1/9\simeq 0.1111$ sufficiently close to $x=0.11$. According
to our theory, it is likely that the $x=1/9$ sample corresponds to the
metastable LTO($9,1,2$) superconducting phase, as shown in Fig. \ref{fig8}.
In this case, the sufficient large stripe-stripe distances ($D_{xy}=25.65%
\mathring{A}$ and $D_{z}=17.45\mathring{A}$) indicate that the stripe-stripe
interaction is much weaker in the compound system with the stiffness
criterion $\Theta =2.24\times 10^{-4}\kappa $. As a result, the $x=1/9$
sample has a softer characteristic and shows a relatively larger positive $%
dT_{c}/dP$ value.

\subsection{SmO$_{1-x}$F$_{x}$FeAs}

We now turn to the SmO$_{1-x}$F$_{x}$FeAs compound system with the lattice
constants $a=$ $2.788\mathring{A}$ and $c=8.514\mathring{A}$. In the
previous paper, the analytical results indicated that the optimum doping
occurs at $x=1/6\approx 0.1667$ in SmO$_{1-x}$F$_{x}$FeAs with the square
vortex lattice of LTT2($6,1,2$) phase. \cite{huang0} Immediately after the
discovery of the $T_{c}$ of 55 $K$ in SmO$_{1-x}$F$_{x}$FeAs, the pressure
effects on the superconducting\ of the new compound have been investigated
by Lorenz \textit{et al}. \cite{lorenz} However, unlike LaO$_{1-x}$F$_{x}$%
FeAs, it was shown that the pressure can either suppress or enhance $T_{c}$,
depending on the doping level.
\begin{figure}[tbp]
\begin{center}
\resizebox{1\columnwidth}{!}{
\includegraphics{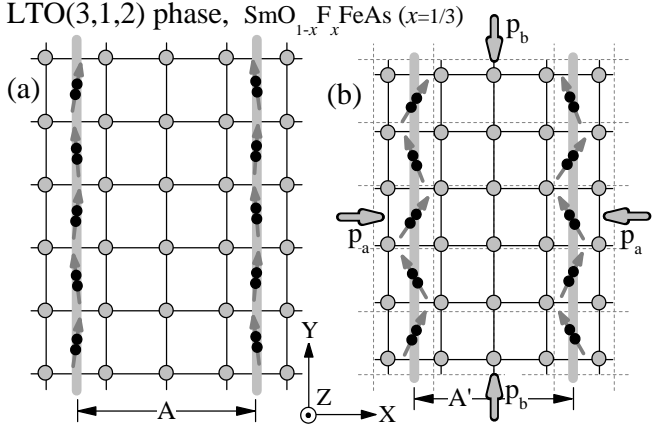}}
\end{center}
\caption{Pressure effects on the overdoped LTO($3,1,2$) superconducting
phase of SmO$_{1-x}$F$_{x}$FeAs ($x=1/3$). (a) Without pressure, (b)
pressure-induced intensive spin fluctuation due to a strong stripe-stripe
interaction.}
\label{fig9}
\end{figure}

According to the present scenario, it is then not a surprise to learn the
doping-dependent pressure effects on $T_{c}$. For the underdoped $x=0.13<1/6$
SmO$_{1-x}$F$_{x}$FeAs sample, the corresponding vortex lattice may be in a
mixed superconducting phase of LTO($7,1,2$) and LTO($8,1,2$) with $%
x=(1/7+1/8)/2$. Apparently, the stripe-stripe distances of the mixed phase
are large enough to support a positive pressure effect [similar to the case
of LTO($9,1,2$) phase for LaO$_{1-x}$F$_{x}$FeAs ($x=0.11$)]. Furthermore,
the mixed vortex phase which exhibits a value of $\Theta =2.77\times
10^{-4}\kappa $ is found harder than the LTO($9,1,2$) phase of LaO$_{1-x}$F$%
_{x}$FeAs ($x=0.11$) with $\Theta =2.24\times 10^{-4}\kappa $. As a
consequence, the LaO$_{1-x}$F$_{x}$FeAs ($x=0.11$) sample should have a
larger $dT_{c}/dP$ value than that of the SmO$_{1-x}$F$_{x}$FeAs ($x=0.13$)
sample, in reasonable agreement with the experiments [$1.2$ K/GPa for LaO$%
_{1-x}$F$_{x}$FeAs ($x=0.11$) \cite{wlu} and $0.9$ K/GPa for SmO$_{1-x}$F$%
_{x}$FeAs ($x=0.13$) \cite{lorenz}].

While for the overdoped $x=0.3>1/6$ SmO$_{1-x}$F$_{x}$FeAs sample,
approximately, the LTO($3,1,2$) and LTO($6,1,1$) superconducting phases are
candidates for the vortex lattices. An example of LTO($3,1,2$) is shown in
Fig. \ref{fig9}, under such circumstances, the stripes (vortex lines) are
very crowd in the superconducting Fe planes with a stripe-stripe distance $%
D_{xy}=8.363\mathring{A}$. It is obvious that the external pressure could
lead to a much more crowded and unstable vortex phase. This in turn greatly
enhance the spin fluctuation and suppress superconductivity, implying a
possibility of \ a negative $dT_{c}/dP\simeq -$ 2.3 K/GPa as indicated in
the experiment of SmO$_{1-x}$F$_{x}$FeAs ($x=0.3$) sample, \cite{lorenz}
having a large negative value of $\Theta =-5.06\times 10^{-4}\kappa .$

\subsection{Uniaxial pressure effects}

The uniaxial pressure effects are markedly different from those of
hydrostatic pressure effects. For example in an single crystal of YBa$_{2}$Cu%
$_{3}$O$_{7-\delta }$, \cite{welp,meingast} the uniaxial pressure dependence
measurements revealed the following pressure derivatives: $%
dT_{c}/dp_{a}=-2.0\pm 0.2$ K/Gpa and $dT_{c}/dp_{b}=+1.9\pm 0.2$ K/Gpa,
where the subscripts ($i=a,b$) denote the corresponding crystallographic
directions. It was shown that the \textbf{a}-axis and \textbf{b-}axis
derivatives are of opposite sign. Note that the anisotropic pressure
dependence of $T_{c}$ along the \textbf{a} and \textbf{b} directions is
still a theoretical challenge.
\begin{figure}[tbp]
\begin{center}
\resizebox{0.9\columnwidth}{!}{
\includegraphics{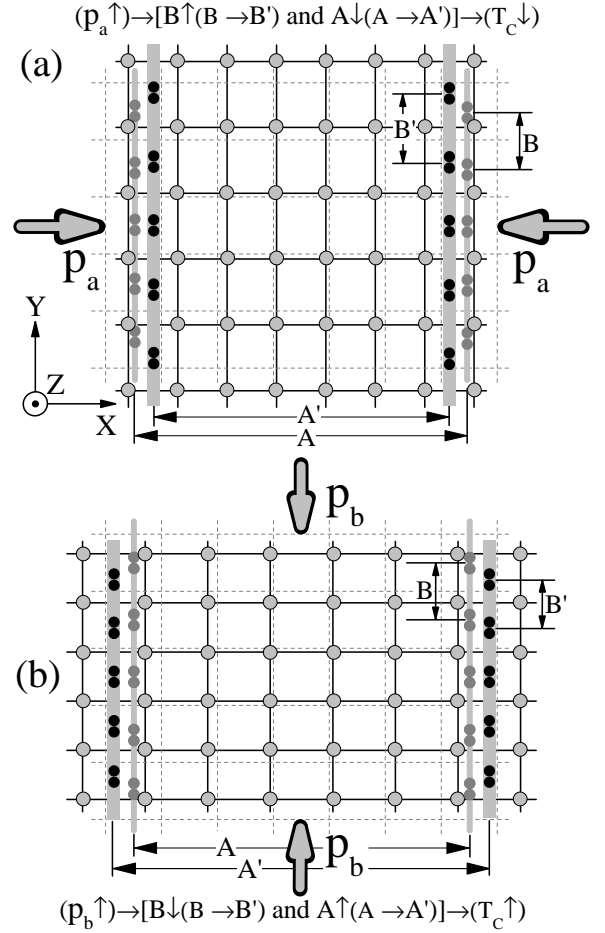}}
\end{center}
\caption{The schematic interpretation of the uniaxial pressure effects. (a)
When the external pressure $p_{a}$ perpendicular to the stripe direction, $%
dT_{c}/dp$ value is negative, (b) when the pressure $p_{b}$ along the stripe
direction, $dT_{c}/dp$ value is positive.}
\label{fig10}
\end{figure}

In this paper we show how the real-space vortex model of Fig. \ref{fig1} can
explain these peculiar results. Furthermore, we argue the existence of the
uniaxial pressure effects in FeAs superconductors. In fact, the uniaxial
pressure dependence in superconductors can be well understood simply by
considering the pressure effects in two special directions : (i) along the
charge stripe (vortex line) direction, and (ii) perpendicular to the stripe
direction. Figure \ref{fig10} illustrates the uniaxial pressure effects in
FeAs superconducting family. When the external pressure is in $\mathbf{a}$%
-axis direction (perpendicular to the stripes), our study reveals that the
pressure affects the vortex lattice at least two factors: shorten the
stripe-stripe spacing while at the same time increase the distance between
cooper pairs and two electrons inside a cooper pair. As discussed above,
both factors are negative for promoting the $T_{c}$, therefore, $dT_{c}/dp$
is expected to be negative for the $\mathbf{a}$-axis pressure. When the
external pressure is in $\mathbf{b}$-axis direction (along the stripes) as
shown in Fig. \ref{fig10}(b), contrary to Fig. \ref{fig10}(a), the
pressure-induced vortex lattice distortions in superconducting Fe planes are
always positive for superconducting, thus a positive $dT_{c}/dp$ may be more
prominent for the $\mathbf{b}$-axis pressure. It should be pointed out that
these results of uniaxial pressure effects are valid for any layered
superconductors.

These results imply that, if pressure is applied only in $\mathbf{a}$-axis
direction, the underdoped LaO$_{1-x}$F$_{x}$FeAs ($x=0.11$) and SmO$_{1-x}$F$%
_{x}$FeAs ($x=0.13$) samples probably have a negative $dT_{c}/dP$ value,
while for the overdoped SmO$_{1-x}$F$_{x}$FeAs ($x=0.3$) sample, the pure $%
\mathbf{b}$-axis pressure may induce a positive $dT_{c}/dp$ pressure effect.

\section{ Concluding remarks}

In conclusion, without Hamiltonian, without wave function, without quantum
field theory, our scenario has provided a beautiful and consistent picture
for describing the high-pressure effects (hydrostatic and uniaxial pressure)
in the newly discovery of the iron-based superconductors. We insist that any
pressure-induced phenomena should share exactly the same physical reason.
The suggested mechanism responsible for the pressure effect is not specific
to the iron-based family and it may also be applicable to other
superconducting materials, including the cuprate superconductors.

\section*{Acknowledgments}

The author would like to thank Dr. Kezhou Xie for many useful suggestions.

\end{document}